\documentclass[aps,prb,twocolumn,showpacs,superscriptaddress]{revtex4-1}
\usepackage{bm}
\usepackage{graphicx}
\usepackage{amsmath}
\usepackage{amssymb}
\usepackage{longtable}
\usepackage{color}
\usepackage[usenames,dvipsnames]{xcolor}
\usepackage[colorlinks=true,linkcolor=Maroon,citecolor=OliveGreen,urlcolor=Blue,linktoc=page]{hyperref}

\newcommand{\Tc}{T_{\mathrm{c0}}}
\newcommand{\Hct}{H_{\mathrm{c2}}}

\newcommand{\N}{{\mathfrak{N}}}

\def\XXint#1#2#3{{\setbox0=\hbox{$#1{#2#3}{\int}$ }
\vcenter{\hbox{$#2#3$ }}\kern-.5\wd0}}

\begin{document}

\title{Analysis of the Ghost and Mirror Fields in the Nernst Signal Induced by Superconducting Fluctuations}
\author{A. Glatz}
\affiliation{Materials Science Division, Argonne National Laboratory, 9700 S. Cass
	Avenue, Argonne, Illinois 60639, USA}
\affiliation{Department of Physics, Northern Illinois University, DeKalb, Illinois
	60115, USA}
\author{A. Pourret}
\affiliation{Université Grenoble Alpes, CEA, IRIG, PHELIQS, F-38000 Grenoble, France}
\author{A.A. Varlamov}
\affiliation{CNR-SPIN, c/o DICII-Universitá di Roma Tor Vergata, Via del Politechnico,
	1, 00133 Roma, Italia}

\date{\today}

\begin{abstract}
We present a complete analysis of the Nernst signal due to superconducting fluctuations in a large variety of  superconductors from conventional to unconventional ones. 
A closed analytical expression of the fluctuation contribution to the Nernst signal is obtained in a large range of temperature and magnetic field. 
We apply this expression directly to experimental measurements of the Nernst signal in Nb$_x$Si$_{1-x}$ thin films and a URu$_2$Si$_2$ superconductors.
Both magnetic field and temperature dependence of the available data are fitted with very good accuracy using only two fitting parameters, the superconducting temperature $\Tc$ and the upper critical field $\Hct(0)$. 
The obtained values agree very well with experimentally obtained values. 
We also extract the ghost lines (maximum of the Nernst signal for constant temperature or magnetic field) from the complete expression and also compare it to several experimentally obtained curves.
Our approach predicts a linear temperature dependence for the ghost critical field well above $\Tc$. 
Within the errors of the experimental data, this linearity is indeed observed in many superconductors far from $\Tc$.
\end{abstract}


\maketitle

\section{Introduction}
In a superconductor above its critical temperature, $\Tc$,  global superconducting coherence vanishes, leaving behind droplets of short lived Cooper pairs. 
Superconducting fluctuations, discovered in the late 1960s, have constituted an important research area in superconductivity as they are manifested in a variety of phenomena. 
Today their investigation has emerged as a powerful tool for quantifying material parameters of new superconductors.  
In this regard, the observation of a giant Nernst signal (three orders of magnitude larger than the value of the corresponding coefficient in typical metals) over a wide range of temperatures and magnetic fields attracted great attention of the superconductivity community and caused lively theoretical discussions~\cite{Uss1,SSVG09,Michaeli2,LNV11,Reizer11,SSVG11}. 
Important milestones were its discovery in underdoped phases of high-temperature superconductors~\cite{XOWKU00}, later in the normal phase of conventional superconductors~\cite{Aubin1,Aubin2}, in normal phases of overdoped, optimally doped, and in the underdoped superconductors La$_{1.8-x}$Eu$_{0.2}$S$r_{x}$CuO$_{4}$, Pr$_{2-x}$Ce$_{2}$CuO$_{4}$~\cite{Tafti14}, and, finally, the observation of the colossal thermo-magnetic response in the exotic heavy Fermion superconductor URu$_{2}$Si$_{2}$~\cite{Yamashita15}.
Today, it is commonly agreed that this effect is related to superconducting fluctuations, and its profound relationship to the fluctuation magnetization is well established~\cite{Behnia,Tafti14,RMP18}.

One of the characteristic features of the fluctuations induced Nersnt signal is its non-monotonous dependence on applied magnetic fields. 
The latter follows from a very generic heuristic arguments: 
the fluctuations induced Nersnt signal is the response to an applied crossed magnetic field and temperature gradient, $\N^{(\mathrm{fl})}=\beta_{xy}^{(\mathrm{fl})}R_{\square}$, where $\beta_{xy}^{(\mathrm{fl})}$ is the off-diagonal component of the fluctuation induced contribution to the thermoelectric tensor~\cite{LV09} and $R_{\square}$ is the film sheet resistance. 
Hence, it is zero at $H=0$ (where the thermoelectric tensor is diagonal) and it should vanish in very strong fields, which suppress fluctuations~\cite{LV09}.
Indeed, a maximum of the Nernst signal as function of the magnetic field has been widely observed~\cite{Tafti14,Yamashita15,Aubin1,Aubin2}.
The study of the temperature dependence of the field at which the Nernst signal is maximum, the \textit{ghost (critical) field} $H^{\ast}(T)$, acquired special significance for HTS compounds, since the authors of Refs.~\onlinecite{Tafti14,Yamashita15} have proposed to use it for the precise determination of the second critical field $\Hct(0)$, often inaccessible for direct measurements because of its huge value.

\section{The issue of the Ghost Field Temperature Dependence}
\begin{figure*}[bth]
	\includegraphics[width=0.8\linewidth]{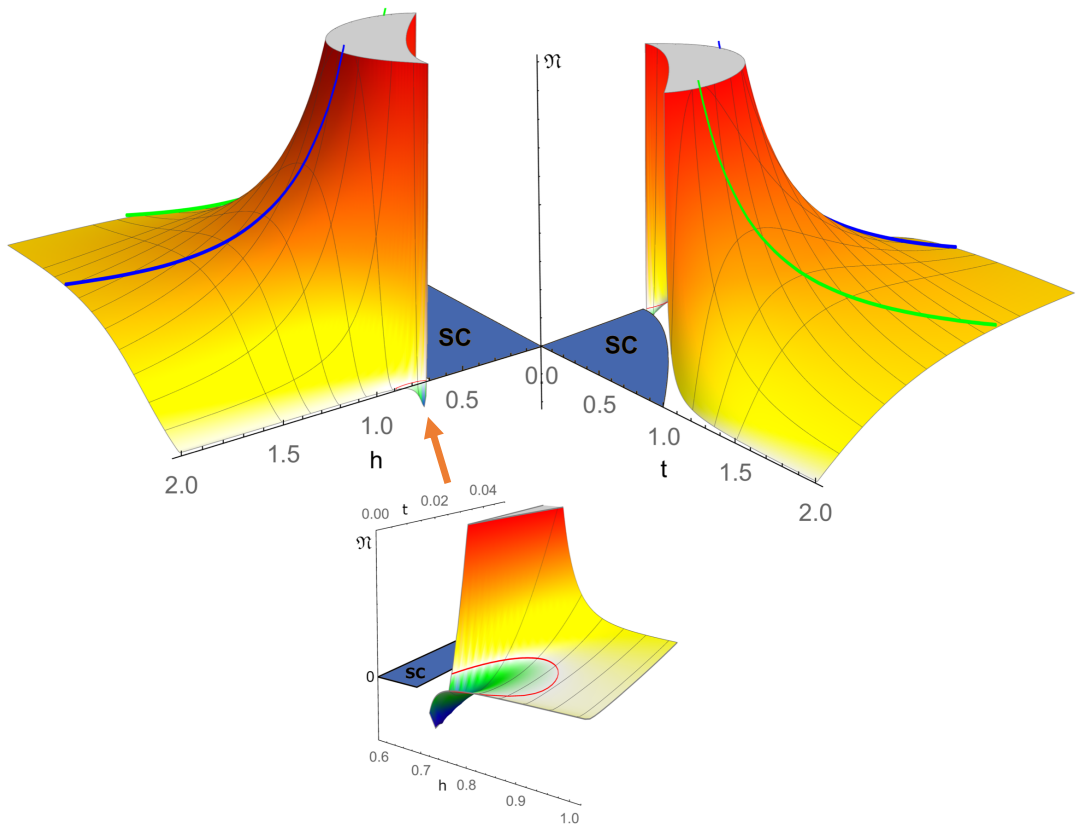} 
	\caption{The magnetic field and temperature dependence of the fluctuation part
		of the Nernst coefficient. \textit{top-left:} A view on the $t=0$ plane with the
		ghost temperature line in blue (light gray) indicating the maximum of the
		Nernst coefficient for constant $h$. \textit{top-right:} A view on the $h=0$ plane with ghost field line in green, indicating the maximum of the Nernst signal for constant fields.
			\textit{bottom:} Zoom on to the quantum fluctuations (QF) region at $t=0$ close to $h=h_{c2}$.
	The red (dark gray) line indicates the contour, where the Nernst signal is zero. In a very small area of the QF region, the Nernst coefficient becomes negative for $t\lesssim 0.02$ and $\tilde{h}\lesssim 0.15$ (see text).}
	\label{fig:nernst} 
\end{figure*}

The analysis of the experimental data obtained from the HTS compound Pr$_{2-x}$Ce$_{2}$CuO$_{4}$ led the authors of Ref.~\cite{Tafti14} to the hypothesis that the temperature dependence of the ``ghost critical field'' is described by the expression: 
\begin{equation}
H^{\ast}_{\mathrm{exp}}(T)=\Hct(0)\ln\frac{T}{\Tc}\,.
\label{eq:ghosttail}
\end{equation}
The prefactor in front of the logarithm with $\Hct(0)$ was determined by observation that $\Hct(0)$ is the only empirical parameter that characterizes the strength of superconductivity. 
The authors stated that ``the characteristic field scale encoded in superconducting fluctuations above $T_{\mathrm{c}}$'', is equal to the field needed to kill superconductivity at $T=0\,K$ and we share this motivation. 
The argument, justifying the logarithmic dependence of $H^{\ast}$ on temperature was based on the statement, that the maximum of the Nernst signal should correspond to the field, where the magnetic length of a fluctuation Cooper pair, $L_{H}$, becomes equal to its ``size''. 
We agree with the latter: 
in terms of the qualitative picture of superconducting fluctuations, one can see how moving along the $\Hct(T)$ line the Ginzburg-Landau long wave-length scenario gradually transforms into the precursor of an Abrikosov vortex lattice: 
a set of clusters of rotating fluctuation Cooper pairs (FCP) in magnetic field, which are relatively small (of size $\sim\xi_{\mathrm{BCS}}$)~\cite{GVV11EPL,RMP18}. 
Yet, in order to practically apply this correct ideological idea, the authors of Refs.~\onlinecite{Aubin1,Aubin2,Tafti14,Taillefer,Yamashita15} extrapolate the Ginzburg-Landau (GL) expression for the FCP coherence length $\xi_{\mathrm{FCP}}\left(T\right)=\xi_{\mathrm{GL}}\left(T\right)\sim\xi_{\mathrm{BCS}}/\sqrt{\ln\frac{T}{\Tc}}$, obtained with the assumption of closeness to the critical temperature~\cite{LV09}, to the region of high temperatures $T\gg \Tc$. 
Indeed, this procedure leads to Eq.~\eqref{eq:ghosttail}. 

However, at this point we need to stress that there is no theoretical justification for such an extrapolation procedure. 
Moreover, it leads to the obviously incorrect conclusion that at high temperatures, the size of FCPs becomes much less than $\xi_{\mathrm{BCS}}$. 
The correlation length $\xi_{\mathrm{FCP}}\left(T\right)$, identified with the fluctuation Cooper pair ``size'', should be determined from the pole of the two-particle Green function, or, \textit{idem}, of the fluctuation propagator~\cite{LV09}. 
For arbitrary temperatures and magnetic fields in impure superconductor, the general form of the latter is: 
\begin{eqnarray}
& L_{n}^{\left(R\right)-1}(-i\omega,q_{z}^{2})\!=\!\label{propagator-1}\\ \label{eq:progen}
& \!-\rho_{e}\left[\ln\frac{T}{\Tc}\!+\!\psi\left(\frac{1}{2}\!+\!\frac{\!-i\omega\!+\!\Omega_{H}(n\!+\!\frac{1}{2})\!+\!Dq_{z}^{2}}{4\pi T}\right)\!-\!\psi\left(\frac{1}{2}\right)\right]\,.\nonumber 
\end{eqnarray}
Close to the critical temperature, where $\ln\frac{T}{\Tc}\approx\frac{T-\Tc}{\Tc}=\epsilon\ll 1$, and in zero magnetic field  it takes the standard form of the diffusive mode, after expansion of the $\psi$-function:
\begin{equation}
L(0,q^2)=-\frac{1}{\rho_{e}}\left(\epsilon+\frac{\pi Dq^{2}}{8T}\right)^{-1}\,.
\label{eq:proclose}
\end{equation}
Analyzing the pole of this expression, $L^{-1}(0,q_0^2)=0$, one indeed obtains $\xi_{\mathrm{FCP}}\left(T\to \Tc\right)\sim q_{0}^{-1}\sim\xi_{\mathrm{GL}}\left(\epsilon\right)\sim\xi_{\mathrm{BCS}}/\sqrt{\epsilon}$.
Yet, far from the critical temperature, with the assumption that $\ln\frac{T}{\Tc}\gg 1$, the $\psi$-function in Eq.~\eqref{eq:progen} with large argument, should be replaced by its asymptotic logarithmic expression and one obtains 
\begin{equation}
L(0,q^{2})=-\frac{1}{\rho_{e}}\ln^{-1}\left(\frac{Dq^{2}}{4\pi \Tc}\right).
\label{eq:profar}
\end{equation}
The pole of this expression is given by $q_{0}^{-1}\sim\sqrt{\frac{4\pi \Tc}{D}}$ resulting in $\xi_{\mathrm{FCP}}\left(T\gg \Tc\right)\sim q_{0}^{-1}\sim\xi_{\mathrm{BCS}}$.
Hence, the qualitative argumentation justifying Eq.~\eqref{eq:ghosttail} is unfounded.

In Ref.~\cite{Kavokin15} the authors looked for an analytical expression for the ghost field by
proposing a scaling arguments based on the general expression for fluctuations induced Nersnt signal (see Refs.~\cite{SSVG09,Sthesis,RMP18}), valid in a wide range of temperatures and magnetic fields.
It was noticed that the magnetic field enters only normalized by temperature, while the latter also appears in the theory as parameter $\ln\left(T/\Tc\right)$. 
This observation allowed them to obtain the following expression for the ghost field, which is very different from Eq.~\eqref{eq:ghosttail}: 
\begin{equation}
H_{\mathrm{KV}}^{\ast}(T)=\Hct(0)\left(\frac{T}{\Tc}\right)\varphi\left(\ln\frac{T}{\Tc}\right)\,,
\label{eq:ghost}
\end{equation}
where $\varphi(x)$ is some smooth function which satisfies the condition $\varphi(0)=0$. 
It is easy to see that Eq.~\eqref{eq:ghost} coincides with Eq.~\eqref{eq:ghosttail} only in the very particular case of $\varphi(x)=x\exp(-x)$.

Due to the extremely cumbersome nature of the general expression for the fluctuations induced Nersnt signal, none of the authors of Refs.~\onlinecite{SSVG09,Michaeli2,Kavokin15} succeeded obtaining an analytical expression for the temperature dependence of the ghost field valid far from the critical temperature. 
Yet, simple equating of the asymptotic expressions valid at low fields and high temperatures $\ln t \gtrsim  1, \, h \ll 1 $ (see Table~\ref{tab:asym}, domain \texttt{VIII}) 
\begin{equation}
\N^{(\mathrm{fl})}\!\left(T,H\right)\!\sim \!\left(\frac{\xi_{\mathrm{BCS}}^{2}}{L_{\Hct}^{2}}\right)\left(\frac{H}{\Hct(0)}\right)\!\left(\frac{\Tc}{T}\right)\!\ln^{-1}\!\frac{T}{\Tc}
\label{eq:Nlow}
\end{equation}
(here $L_{\Hct}^{2}=\frac{c}{2e\Hct(0)}\sim\xi_{\mathrm{BCS}}^{2}$)
and that one valid for high fields $h \gg \max \{ 1,t \}$ (see Table~\ref{tab:asym}, domain \texttt{IX}) 
\begin{equation}
\mathfrak{N}^{(\mathrm{fl})}\!\left(T,H\right)\!\sim\!\left(\frac{L_{H_{c2}}^{2}}{\xi_{\mathrm{BCS}}^{2}}\right)\!\!\left(\!\frac{T}{\Tc}\right)\left(\frac{\Hct(0)}{H}\!\right)\!\ln^{-1}\!\frac{H}{\Hct(0)}
\label{eq:Nhigh}
\end{equation}
leads to the conclusion that at sufficiently high temperatures ($T \gtrsim \Tc$) the ghost critical field should grow as function of temperature almost linearly (with logarithmic accuracy): 
\begin{equation}
H^{\ast}(T)\sim \Hct(0)\left(\frac{T}{\Tc}\right)\,.
\label{eq:ghosteval}
\end{equation}

\begin{table*}[tbh]
	{ \ %
	\begingroup
\setlength{\tabcolsep}{12pt} 
\renewcommand{\arraystretch}{1.8} 

		\begin{tabular}{|l|c|l|c|}
			\hline 
			\textbf{domain}  & $t$ \textbf{and} $h$ \textbf{range} & \textbf{description} & $\pi\N^{(\mathrm{fl})}/\N_0$\tabularnewline
			\hline \hline
			\texttt{I} & $h = 0$, $\epsilon \ll 1$ & zero field, near $\Tc$ & $\frac{2eH\xi_{\mathrm{GL}}^{2}(T)}{3c}=\frac{h}{3\epsilon}$\tabularnewline
			\hline 
			\texttt{II}  & $\epsilon \ll h \ll 1$ & near $\Tc$, above the mirror reflected $\Hct$-line & $1-(\ln2)/2$\tabularnewline
			\hline 
			\texttt{III} & $h - \Hct(t) \ll 1$, $\epsilon \ll 1$ & near $\Hct$-line & $\frac{1}{\epsilon+h}$\tabularnewline
			\hline 
			\texttt{IV} &$t \ll \widetilde{h}$ & region of quantum fluctuations & $-\frac{2\gamma_{\mathrm{E}}}{9}\frac{t}{{\tilde{h}}}$\tabularnewline
			\hline 
			\texttt{V} & $ t^{2}/\ln (1/t) \ll \tilde{h}\ll t\ll 1$ & quantum-to-classical & $\ln\frac{t}{\tilde{h}}$\tabularnewline
			\hline 
			\texttt{VI} & $\tilde{h} \ll t^{2}/\ln (1/t)\ll 1$ & classical, near $\Hct(t \ll	1)$ & $\frac{8\gamma_{\mathrm{E}}^{2}}{3}\frac{t^{2}}{\tilde{h}(t)}$\tabularnewline
			\hline 
			\texttt{VII} & $t^{2}/\ln (1/t) \lesssim \widetilde{h}(t) \ll 1  $ & classical, intermediate fields & $\frac{1}{\tilde{h}(t)}\left[ 1 + \frac{2\Hct(t)}{\pi^{2}t}\frac{\psi^{\prime \prime }(\frac{1}{2} + \frac{2\Hct(t)}{\pi^{2}t})}{\psi^{\prime }(\frac{1}{2} + \frac{2\Hct(t)}{\pi^{2}t})}\right]$\tabularnewline
			\hline 
			\texttt{VIII} & $\ln t \gtrsim 1, h \ll t$ & high temperatures   & $\frac{2}{3\pi^{2}}\frac{h}{t\ln t}$\tabularnewline
			\hline 
			\texttt{IX} & $h \gg \max \{1, t\}$ & high magnetic fields & $\frac{\pi^{2}}{48}\frac{t}{h\ln{h}}$\tabularnewline
			\hline 
	\end{tabular}
	\endgroup
	} \caption{Asymptotic expressions (obtained in Ref.~\onlinecite{SSVG09}) for fluctuation
		corrections to the Nernst signal in different domains of the phase
		diagram (see Fig.~\ref{fig:phases}). 
		Where $\N_0\equiv \frac{\pi\hbar}{k_{B}R_{\Box}}$,
		 $\tilde{h}=\frac{h-h_{\mathrm{c2}}(t)}{h_{c2}(t)}\ll1$,
		and $h_{c2}(0)=\frac{\Hct(0)}{\widetilde{H}_{\mathrm{c2}}(0)}=\frac{\pi^{2}}{8\gamma_{\mathrm{E}}}=0.69$
		(see text).}
	\label{tab:asym} 
\end{table*}

In Ref.~\onlinecite{RMP18}, a general computational approach to the description of fluctuation phenomena in superconductors, valid in the whole phase diagram, \textit{numerical fluctuoscopy}, was presented. 
In the  following we will apply this method for the determination of the true temperature dependence of the ghost field in the Nernst signal and its comparison with experimental data.

\section{Consistent Derivation of the Ghost Field}

\subsection{Theoretical Foundation: fluctuations induced Nersnt Signal}

\begin{figure}[h]
	\includegraphics[width=0.9\columnwidth]{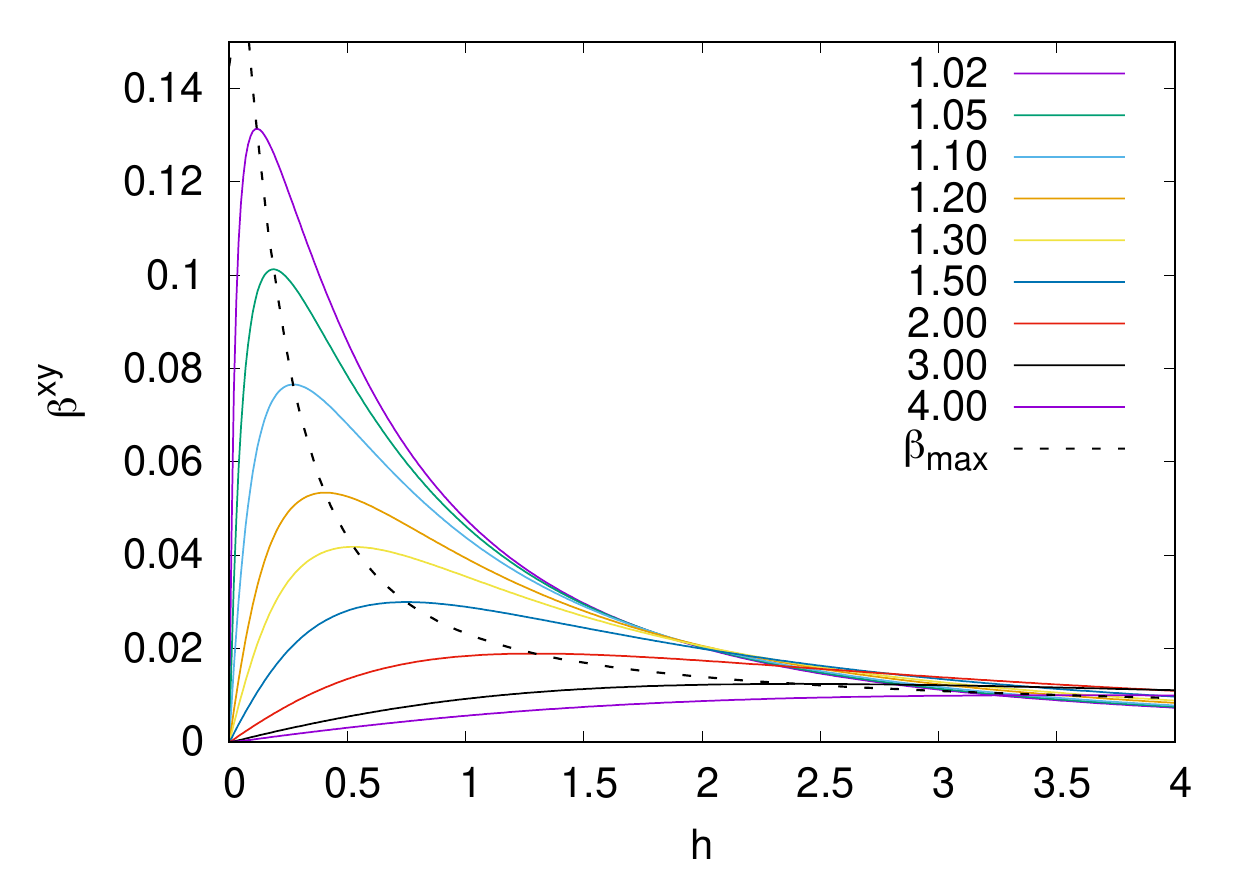} \\	\includegraphics[width=0.9\columnwidth]{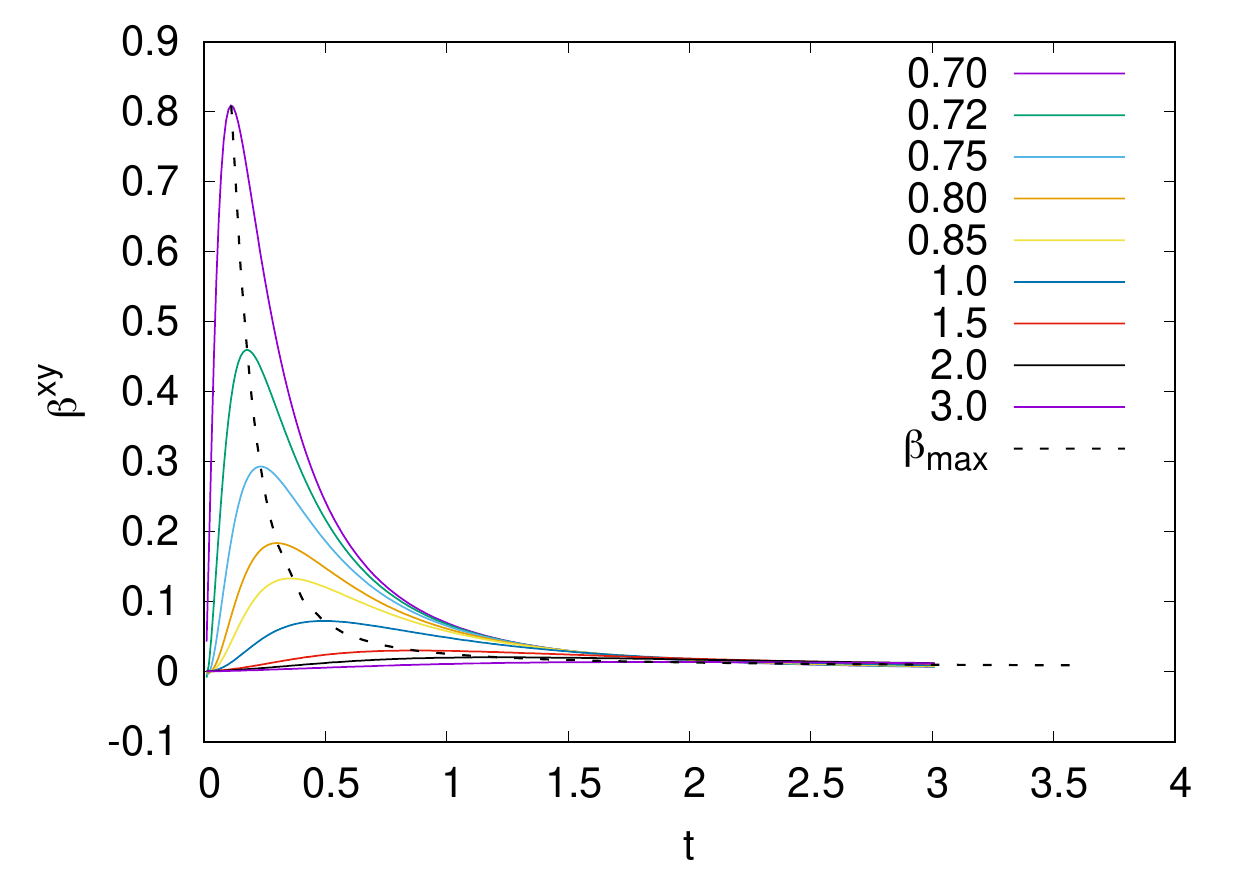}
	\caption{fluctuations induced Nersnt signal as  function of (a) magnetic field at constant temperatures (above $\Tc$, i.e., $t>1$), where the reduced temperatures are indicated in the legend and the curve of maxima (dashed) as function of temperature and  (b) as function of temperature at constant
		fields (above $h_{c2}(0)$), indicated in the legend, with dashed maxima curve as function of field.}
	\label{fig:nernstdet} 
\end{figure}

The general expression for the fluctuation contribution to the Nernst signal of 2D superconductors, valid beyond the line $\Hct(t)$, can be presented in the form suitable for the numerical analysis as~\cite{SSVG09,Sthesis,RMP18}: 
\begin{widetext}
	\small{
	\begin{eqnarray}
		\N^{(\mathrm{fl})} & \!=\! & \frac{\N_0}{8\pi}\!\left[\sum_{m=0}^{M_{t}}(m\!+\!1)\!\!\sum_{k=-\infty}^{\infty}\!\left\{ \!\left(\!\frac{\eta(2m\!+\!3)\!+\!|k|}{\mathcal{E}_{m}}\!+\!\frac{\eta(2m\!+\!1)\!+\!|k|}{\mathcal{E}_{m\!+\!1}}\right)\left(\mathcal{E}_{m}^{\prime}\!-\!\mathcal{E}_{m\!+\!1}^{\prime}\right)\!+\!2\eta\left[\eta\left(2m\!+\!1\right)\!+\!|k|\right]\frac{\mathcal{E}_{m}^{\prime\prime}}{\mathcal{E}_{m}}\!+\!2\eta\left[\eta(2m\!+\!3)\!+\!|k|\right]\frac{\mathcal{E}_{m\!+\!1}^{\prime\prime}}{\mathcal{E}_{m\!+\!1}}\right\} \right.\nonumber\\
		&  & +4\pi^{2}\sum_{m=0}^{M_{t}}(m+1)\int_{-\infty}^{\infty}\frac{dx}{\sinh^{2}\pi x}\left\{ \frac{\eta\mathop{{\rm Im}}\mathcal{E}_{m}\mathop{{\rm Im}}\left(\mathcal{E}_{m}+\mathcal{E}_{m+1}\right)+\left[\eta(m+1/2)\mathop{{\rm Im}}\mathcal{E}_{m}+\frac{x}{2}\mathop{{\rm Re}}\mathcal{E}_{m}\right]\mathop{{\rm Im}}\left(\mathcal{E}_{m+1}+2\eta\mathcal{E}_{m}^{\prime}-\mathcal{E}_{m}\right)}{|\mathcal{E}_{m}|^{2}}\right.\nonumber\\
		&  & +\frac{\eta\mathop{{\rm Im}}\mathcal{E}_{m+1}\mathop{{\rm Im}}\left(\mathcal{E}_{m}+\mathcal{E}_{m+1}\right)+\left[\eta(m+3/2)\mathop{{\rm Im}}\mathcal{E}_{m+1}+\frac{x}{2}\mathop{{\rm Re}}\mathcal{E}_{m+1}\right]\mathop{{\rm Im}}\left(\mathcal{E}_{m+1}+2\eta\mathcal{E}_{m+1}^{\prime}-\mathcal{E}_{m}\right)}{|\mathcal{E}_{m+1}|^{2}}+2x\mathop{{\rm Im}}\ln\frac{\mathcal{E}_{m}}{\mathcal{E}_{m+1}}\label{eq:complete}\\ 
		&  & \left.\left.\!-\!2\frac{\mathop{{\rm Im}}\left(\mathcal{E}_{m}\!+\!\mathcal{E}_{m\!+\!1}\right)\left(\mathop{{\rm Im}}\mathcal{E}_{m}\mathop{{\rm Im}}\mathcal{E}_{m\!+\!1}\!+\!\mathop{{\rm Re}}\mathcal{E}_{m}\mathop{{\rm Re}}\mathcal{E}_{m\!+\!1}\right)}{|\mathcal{E}_{m\!+\!1}|^{2}|\mathcal{E}_{m}|^{2}}\Big[\eta\left(m\!+\!\frac{3}{2}\right)\mathop{{\rm Im}}\mathcal{E}_{m+1}\!-\eta\left(m+\frac{1}{2}\right)\mathop{{\rm Im}}\mathcal{E}_{m}\!+\!\frac{x}{2}\mathop{{\rm Re}}\left(\mathcal{E}_{m\!+\!1}\!-\!\mathcal{E}_{m}\right)\Big]\right\} \right]\nonumber
	\end{eqnarray}}
\end{widetext}
with $\N_0=\frac{ek_BR_{\square}}{\hbar}$.
Here the function 
\begin{equation}
\mathcal{E}_{m}\equiv\mathcal{E}_{m}(t,h,|k|)\!=\!\ln t+\psi\left[\frac{|k|+1}{2}\!+\!\eta\left(m\!+\!\frac{1}{2}\right)\right]-\psi\left(\frac{1}{2}\right)\label{eq:Em}
\end{equation}
is the denominator of the above mentioned fluctuation propagator.
Its derivatives with respect to the argument $x$ are related to polygamma functions: 
\begin{align}
\mathcal{E}_{m}^{(n)}(t,h,x) & \equiv\frac{\partial^{n}}{\partial x^{n}}\mathcal{E}_{m}(t,h,x)\nonumber \\
& =2^{-n}\psi^{(n)}\!\left[\frac{1+x}{2}+\eta\left(m+\frac{1}{2}\right)\right]\,.\label{eq:dEm}
\end{align}
We use here the convenient combination $\eta=\!\frac{4h}{\pi^{2}t}$ of the dimensionless temperature $t=\frac{T}{\Tc}$ and magnetic field $h=\frac{H}{\widetilde{H}_{\mathrm{c2}}(0)}$. 
The latter is normalized by the value of the second critical field obtained by linear extrapolation of its temperature dependence near $\Tc$:
$\widetilde{H}_{\mathrm{c2}}(0)=\Phi_{0}/\left(2\pi\xi^{2}\right)$, where $\Phi_{0}=\pi c/e$ is the magnetic flux quantum. 
The value of the magnetic field $\widetilde{H}_{\mathrm{c2}}(0)$ is $8\gamma_{\mathrm{E}}/\pi^{2}$ times larger than the Abrikosov's value for the second critical field $\Hct(0)$: 
\begin{equation}
h=\frac{H}{\widetilde{H}_{\mathrm{c2}}(0)}=\frac{\pi^{2}}{8\gamma_{\mathrm{E}}}\frac{H}{\Hct(0)}=0.69\frac{H}{\Hct(0)}\,.\label{eq:h}
\end{equation}
In analogy to $\epsilon$, which measures the closeness to $\Tc$ in zero field, we introduce $\tilde{h}(t)=\frac{h-h_{c2}(t)}{h_{c2}(t)}$, where $\tilde{h}(0)$ measures the closeness to the true critical field at zero temperature.
Despite the apparent divergence of Eq.~\eqref{eq:complete} (we introduced the natural upper limit of the summation over Landau levels $M_{t}\sim(T_{c0}\tau)^{-1}$, with $\tau$ being the electron elastic scattering time) it in fact converges due to intricate cancellations in two divergent orders of the transport (Kubo) and magnetization current fluctuation contributions (see Refs.~\onlinecite{SSVG09,YNO64,Kavokin15}). 
This can be verified by expanding all functions $\mathcal{E}_{m}^{(n)}(t,h,x)$ and their derivatives in Eq.~\eqref{eq:Em} over Landau level differences in the limit of large numbers. 
Hence, the result of summations does not depends on the cut-off parameter. 
This fact is also confirmed by the direct numerical evaluation.

\subsection{Numerical Analysis of the fluctuations induced Nersnt Signal}

\begin{figure*}[htb]
	\includegraphics[width=0.8\linewidth]{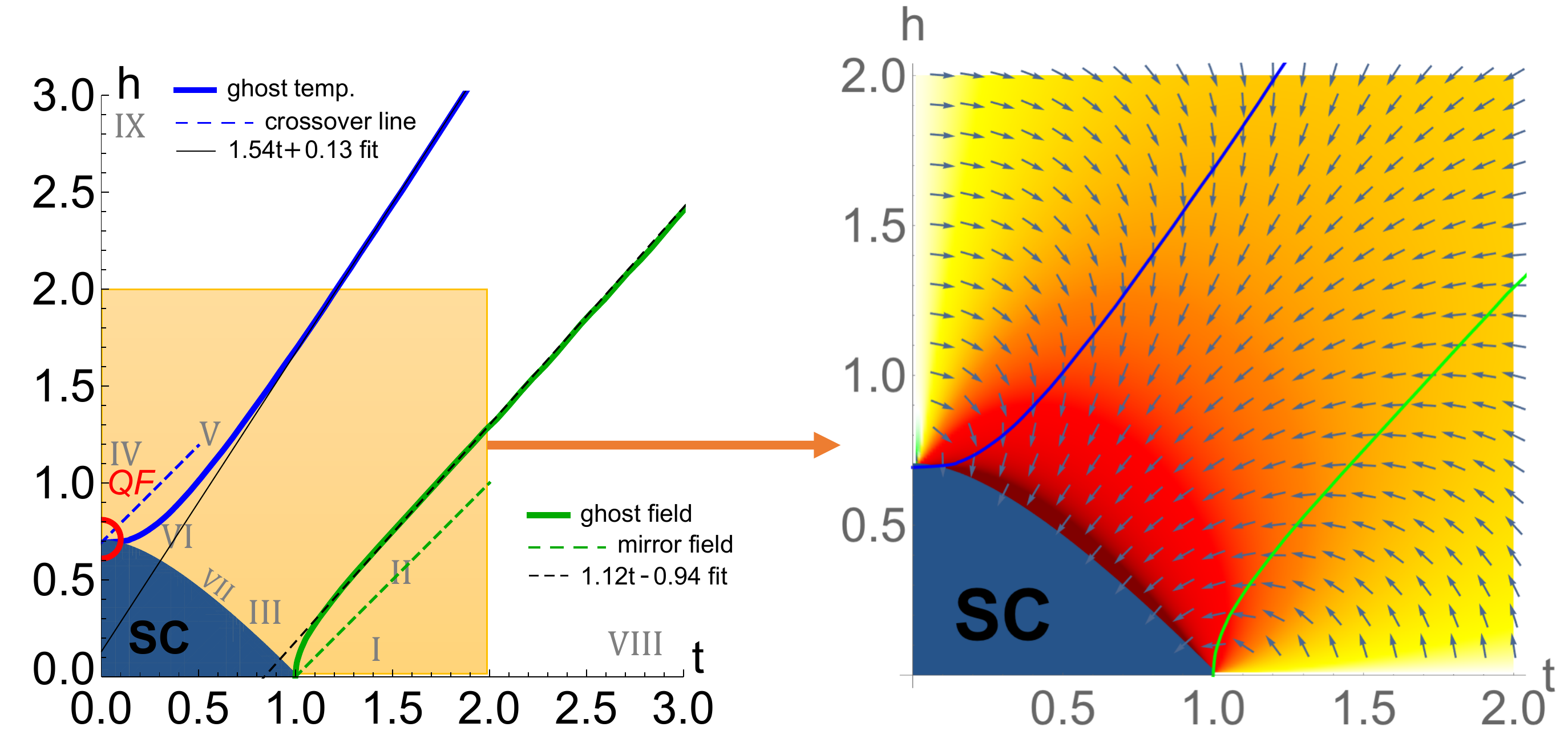} \caption{\textit{Left:} Phase diagram with the lines of the BCS second critical field $h_{c2}(t)$, the ghost field $h^{*}(t)$, the ghost temperature $t^{*}(h)$, the mirror field $h^{(m)}(t)$, and the crossover line from quantum to thermal fluctuations $t^{(\text{qt})}(h)$. 
	The regions of qualitatively different asymptotic behavior is indicated by roman numbers, which are explained in table~\ref{tab:asym}. 
	The region of quantum fluctuations is marked by ``QF'' -- in this region the Nernst coefficient becomes negative.
	The shaded region is enlarged on the \textit{right}, which shows both ghost lines with a density plot  of the Nernst signal.
	In addition the $(t,h)$-gradient of $\N^{(fl)}$ is indicated by a vector field. 
	The ghost lines follow the vertical and horizontal gradients, respectively.}
	\label{fig:phases} 
\end{figure*}

The fluctuation contribution to the Nernst signal in the whole phase diagram beyond the line $\Hct(t)$ as surface plot in accordance to Eq.~\eqref{eq:complete} is presented in Fig.~\ref{fig:nernst}. Fig.~\ref{fig:nernstdet} shows selected isomagnetic and isothermal cuts of this surface plot.
The asymptotic expressions for the Nernst signal in different domains of the phase diagram
are summarized in table~\ref{tab:asym} and the corresponding domains are indicated
in Fig.~\ref{fig:phases}.

Close to the critical temperature $\Tc, $ (domains~\texttt{I}-\texttt{III}) where fluctuations have Ginzburg-Landau thermal character, the Nernst signal is positive and grows in magnitude approaching the transition line ($h-\Hct(t)\ll 1$). 
The ``mirror field'',  $h^{\left(m\right)}(t)=t-1$, separates the linear and non-linear regimes in the magnetic field dependence of the Nernst signal.~\footnote{The notion of ``mirror-symmetric field'' was introduced by Buzdin et al. in Ref.~\onlinecite{BD97} in the study of the nonlinear fluctuation magnetization close to the critical temperature and arbitrary magnetic fields. 
It has the meaning of a crossover line, which separates linear and nonlinear regimes of the fluctuation magnetization, conductivity, and other characteristics~\cite{LV09} in the phase diagram. 
Well before, the exact same meaning of the crossover line between linear and non-linear regimes in fluctuation conductivity was investigated by A. Kapitulnik et al.~\cite{Kapitulnik85}, yet calling it the ``ghost field''. 
In 2000s, after the discovery of the non-monotonous behavior of the fluctuations induced Nersnt signal as function of magnetic field~\cite{Aubin1,Aubin2}, the name ``ghost field' was attributed to the value of magnetic field corresponding to its maximum.  
Following these works and others~\cite{Tafti14} the term ``ghost field'' was henceforth used in the latter sense. 
It is clear, that the ``mirror field'' is always smaller the ``ghost field'' $h^{\ast}(t)$, as demonstrated in Fig.~\ref{fig:NbSi_ghost}.}

The isothermal Nernst signal graphs in Fig.~\ref{fig:nernstdet}a, show the well known non-monotonous behavior of the Nernst signal for temperatures above $\Tc$. 
The dashed line connects the maxima for various fixed values of temperature, indicating the ``ghost field temperature dependence'', which at sufficiently high temperatures is well described by the expression 
\begin{equation}
h^{\ast}(t)\approx 1.12 \left(t-0.84\right)\,,
\label{eq:GF_linfit}
\end{equation}
fit shown in Fig.~\ref{fig:phases}.
One can see that its linear dependence on temperature corresponds to our qualitative picture above and is quite different from the logarithmic law used in Refs.~\onlinecite{Tafti14,Yamashita15}.

Of special interest is the study of the low-temperature regime of fluctuations, close to the upper critical field $\Hct(0)$ (domains~\texttt{IV}--\texttt{VI}). 
Here a crossover line, $t^{(\text{qt})}(h)=\widetilde{h}$, exists, which separates the purely quantum regime at vanishing temperatures (domain~\texttt{IV}) and the region of low temperatures, but where fluctuations already acquire thermal character (domain~\texttt{VI}).
It is interesting, that in the quantum regime the fluctuation contribution to the Nernst signal is negative in a very small $t$-$h$ area, where it depends linear on temperature and diverges as $\tilde{h}^{-1}$ approaching the transition point (see the insert in Fig. 1). 
This change of the sign in the fluctuation thermoelectric response is similar to the negative fluctuation conductivity close to the quantum phase transition in the vicinity of $\Hct(0)$, found in Ref.~\onlinecite{GL01b}. 
These negative values comes from the diffusion coefficient renormalization contribution, which exceeds the positive, but fading away AL term in this region. 
In the quantum-to-classical crossover region (domain~\texttt{V}), the Nernst signal becomes positive and less singular ($\sim\ln\frac{t}{\tilde{h}}$). 
Increasing the temperature, one goes over into the region of thermal fluctuations (domain~\texttt{VI}) and sees that the Nernst signal continue to grow as $\sim t^{2}/\tilde{h}$.

In the isomagnetic Nernst signal plots above the second critical field, shown in Fig.~\ref{fig:nernstdet}b, one sees, similarly to the situation above $\Tc$, that the Nernst signal temperature dependence at fixed fields is non-monotonous and has maximum. 
The line connecting these maxima can be called the ''ghost temperature line'' and it is well described by the linear dependence 
\begin{equation}
t^{\ast}(h)\approx 0.65\left(h-0.13\right)\,,\label{eq:GT_linfit}
\end{equation}
for $h>1$ (see Fig.~\ref{fig:phases}, $1.54t+0.13$ fit).

In the following we will use these insights and complete expression, Eq.~\eqref{eq:complete}, for the Nernst signal to fit experimental data allowing to perform a characterization of the superconducting material. In particular, we can extract the values of $T_{c0}$ and $\Hct(0)$, without using any `artificial' convenience criteria (like half width of the transition region, 90\% of the resistance decay, the temperature where the derivative of resistance is maximal or has an inflection point, etc). 
In a `simplified' version one can just use the ghost field and ghost temperature lines (Eqs.~\eqref{eq:GF_linfit}-\eqref{eq:GT_linfit}) for fitting instead of the non-trivial \textit{fluctuoscopy}, the full fitting procedure of the Nernst signal.

\section{$\text{Nb$_x$Si$_{1-x}$}$ experiments}

In order to verify our theoretical studies, measurements on two stoichiometrically identical samples of Nb$_{x}$Si$_{1-x}$ were performed, labelled samples 1 \& 2 in the following. The Nb concentration, $x$, was fixed at $x=0.15$. 
These amorphous film samples were prepared under ultrahigh vacuum by e-beam coevaporation of Nb and Si with precise control over concentrations and deposited on sapphire substrates. 
Such films typically undergo a metal-insulator transition when $x$ decreases. 

The two samples have different thicknesses, which mostly controls their critical parameters, since the nominal concentration is the same:
Sample 1 (2) was 12.5 (35)~nm thick, its experimental midpoint $\Tc^{\left(\mathrm{exp}\right)}$ was 0.165 (0.380)~K (resistively measured in zero field) and its upper critical field $\Hct^{\left(\mathrm{exp}\right)}(0)$ was 0.36 (0.91)~T. The zero temperature coherence lengths for both samples are 19.7nm and 13nm, respectively

The Nernst coefficient is obtained by measuring the thermoelectric and electric coefficients of both samples in a dilution fridge using a resistive heater, two RuO$_2$ thermometers, and two lateral contacts. Partial data was published in Refs.~[\onlinecite{Aubin1, Aubin2}].
At $T\sim 0.19$K, the d.c. voltage resolution for our setup was 1nV and temperature resolution 0.1mK. 
The results are discussed below.

\section{Nernst signal fluctuoscopy of $\text{Nb$_{0.15}$Si$_{0.85}$}$ and other materials}

\begin{figure}[htb]
	\includegraphics[width=\linewidth]{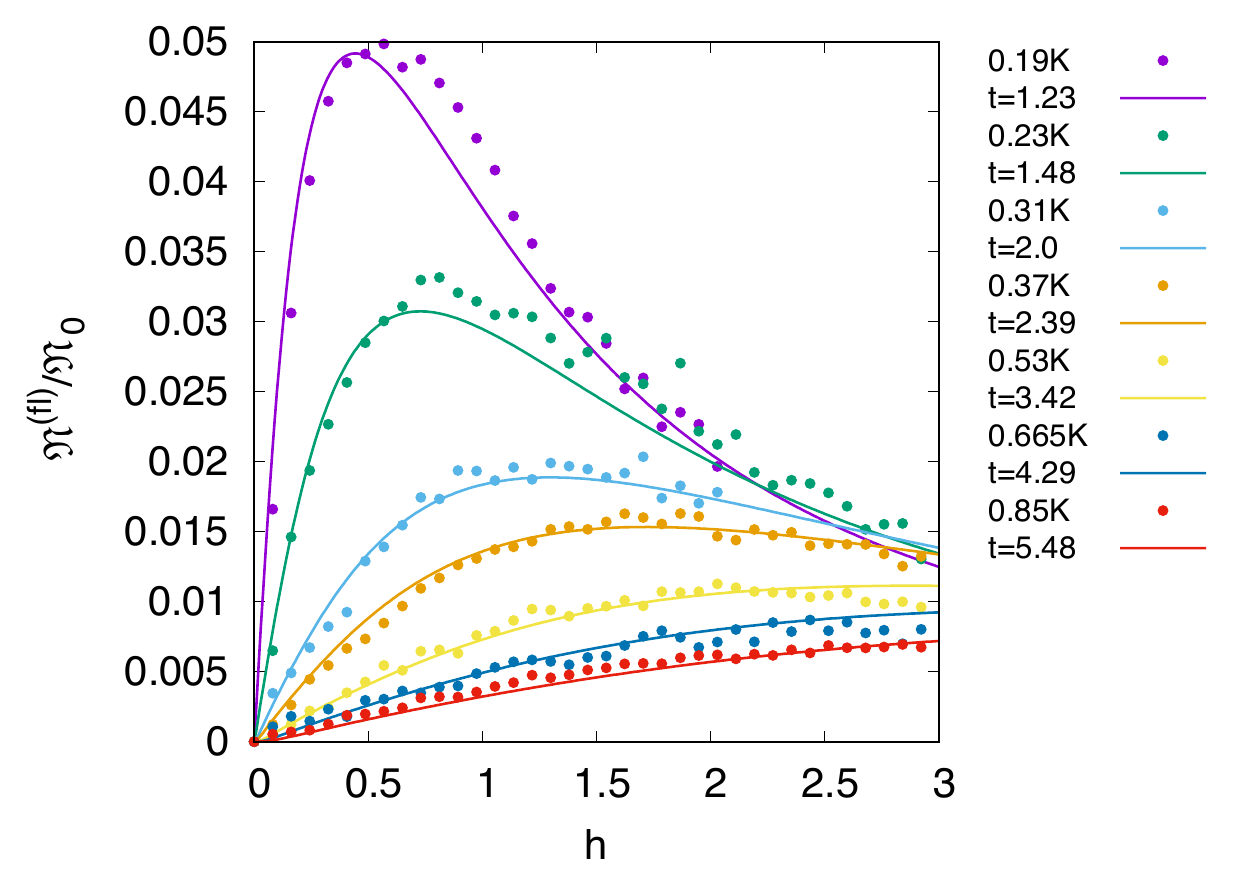} 
	\caption{Fit of the Nernst signal for Nb$_{0.15}$Si$_{0.85}$ (sample 1) to Eq.~\eqref{eq:complete}. The found fitting parameters are $\Tc=0.165$K, $H_{c2}(0)=0.36$T.}
	\label{fig:NbSi_S1} 
\end{figure}

\begin{figure}[htb]
	\includegraphics[width=\linewidth]{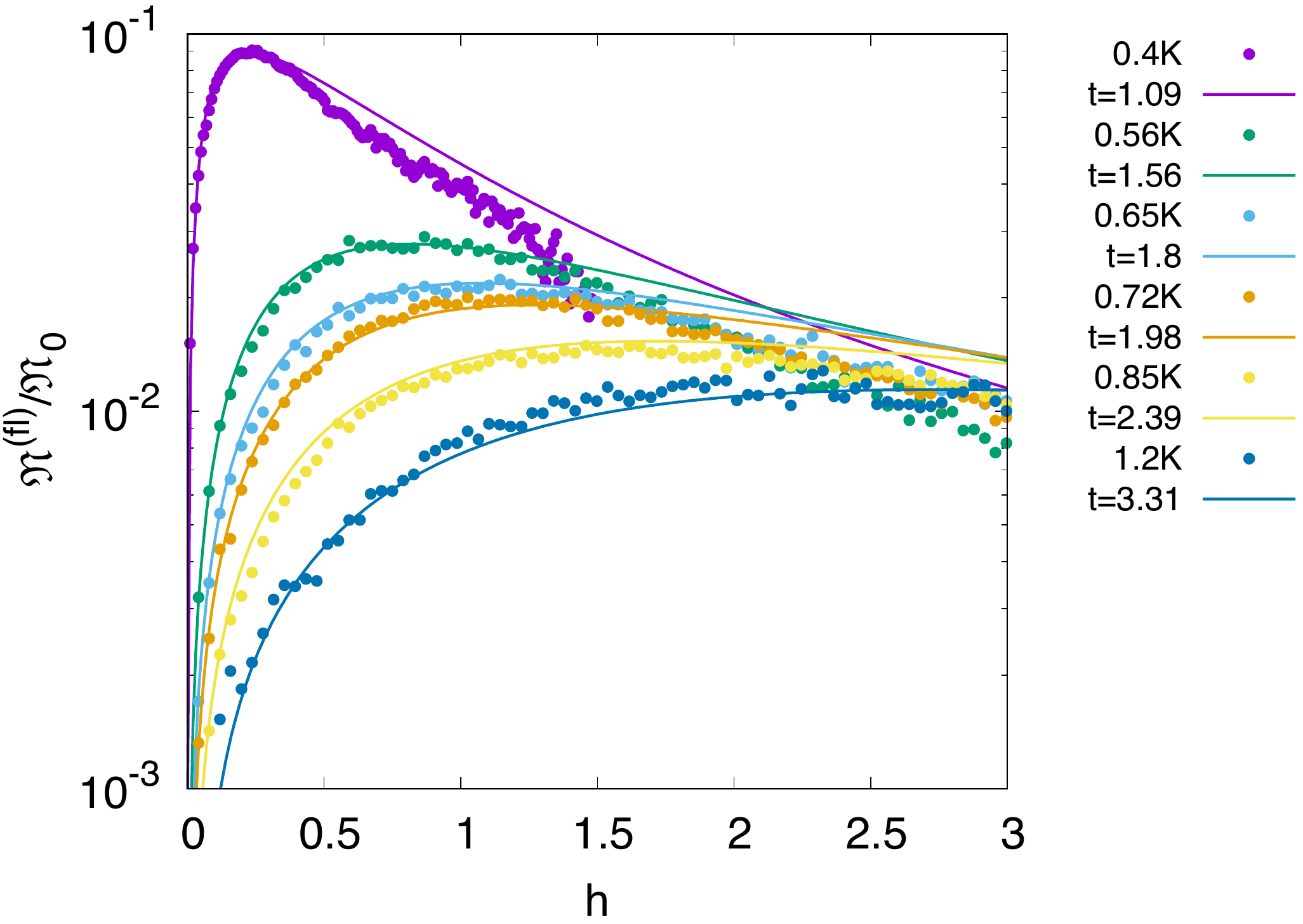} 
	\caption{Fit of the Nernst signal for Nb$_{0.15}$Si$_{0.85}$ (sample 2) to Eq.~\eqref{eq:complete}, shown in half-logarithmic representation. The found fitting parameters are $\Tc=0.36$K, $H_{c2}(0)=0.7$T, which are close to the experimentally estimated values.}
	\label{fig:NbSi_S2} 
\end{figure}

\begin{figure}[htb]
	\includegraphics[width=0.8\linewidth]{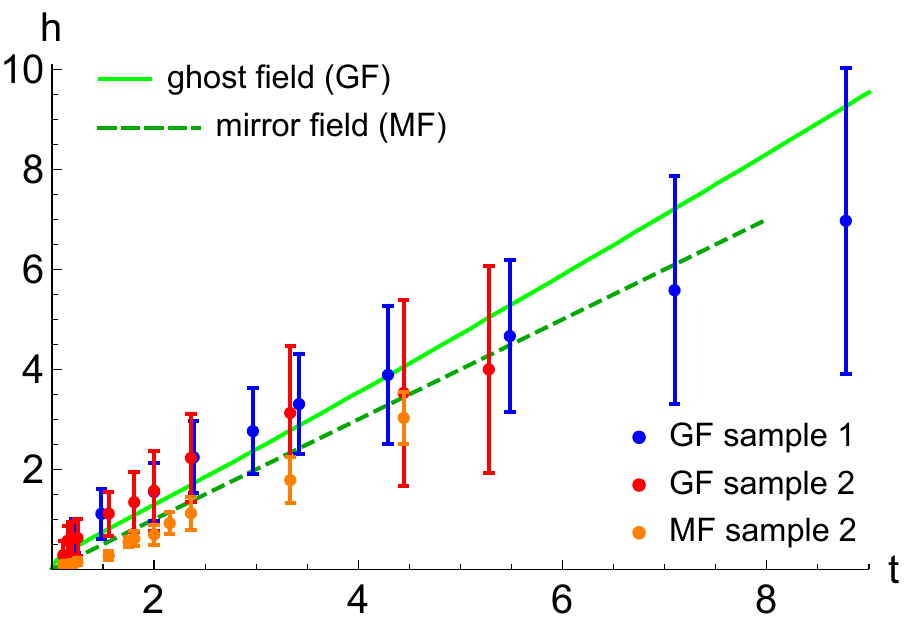} 
	\caption{Numerically evaluated ghost field curve $h^*(t)$ from Eq.~\eqref{eq:complete} in solid green and the mirror field in dashed green. Extracted ghost field from experimental data scaled by fitting parameters with error bars for both samples and the extracted mirror field for sample 2, when the Nernst coefficient starts to deviate from linear behavior. As one can see the error bars for the ghost field become larger for larger temperatures since the maxima become very broad.}
	\label{fig:NbSi_ghost} 
\end{figure}

As already noted, in previous studies the dependence of the fluctuation contribution to the Nernst signal on magnetic field and temperature above the critical one has been fitted~\cite{Behnia} by asymptotic expressions and interpolations between them~\cite{Uss1,SSVG09,Michaeli2,LNV11} with limited accuracy, which also does not allow for a consistent extraction of the ghost lines.
Here we use the general expression Eq.~\eqref{eq:complete} for detailed numerical analysis \& high precision fitting of experimental data in the whole $t$-$h$ plane without the interpolation procedure.

In Figs.~\ref{fig:NbSi_S1} and ~\ref{fig:NbSi_S2} one can see how accurately Eq.~\eqref{eq:complete} fits the experimental data of two Nb$_{0.85}$Si$_{0.15}$ samples using only two fitting parameters: $\Tc$ and $H_{c2}(0)$.
The values of the fitting parameters for sample 1 are $\Tc^{\left(\mathrm{theo}\right)}=0.165 \, K$ and $\Hct^{\left(\mathrm{theo}\right)}(0)=0.36\,T$.

Similarly, the fits of the measurements obtained on sample 2 give the values of critical temperature and second critical field close to their experimentally estimated meanings: $\Tc^{\left(\mathrm{theo}\right)}=0.36\, K$ and $\Hct^{\left(\mathrm{theo}\right)}(0)=0.7\, T$.
The lower values for the critical temperature are in agreement with previous observations, that $\Tc$ is typically overestimated in the experiment~\cite{BG_EPL_12} using traditional convenience methods.

The dependence of the position of maximum in the Nernst signal $\N^{(\mathrm{fl})}\left(h\right)$ versus temperature for  Nb$_{0.85}$Si$_{0.15}$ is shown in the Fig.~\ref{fig:NbSi_ghost}, which demonstrates both the numerically obtained theoretical curve and the values extracted from the experimental data.
One can see that the behavior of $h^{*}\left(t\right)$ obtained from the numerical study of the extremum of Eq.~\eqref{eq:complete} is strongly non-linear close to $\Tc$, but  becomes linear as function of temperature quickly and can be described by Eq.~\eqref{eq:GF_linfit}. 
However, we note that the error bars of the experimentally obtained ghost field become quite large due to the broadness of the maxima, such that the theoretical curve lies well within the error. 
Similar results are obtained for sample 2.

Besides the two Nb$_x$Si$_{1-x}$ samples, we also analyzed several other available Nernst signal measurements using Nernst fluctuoscopy.
\begin{figure}[htb]
	\includegraphics[width=1\linewidth]{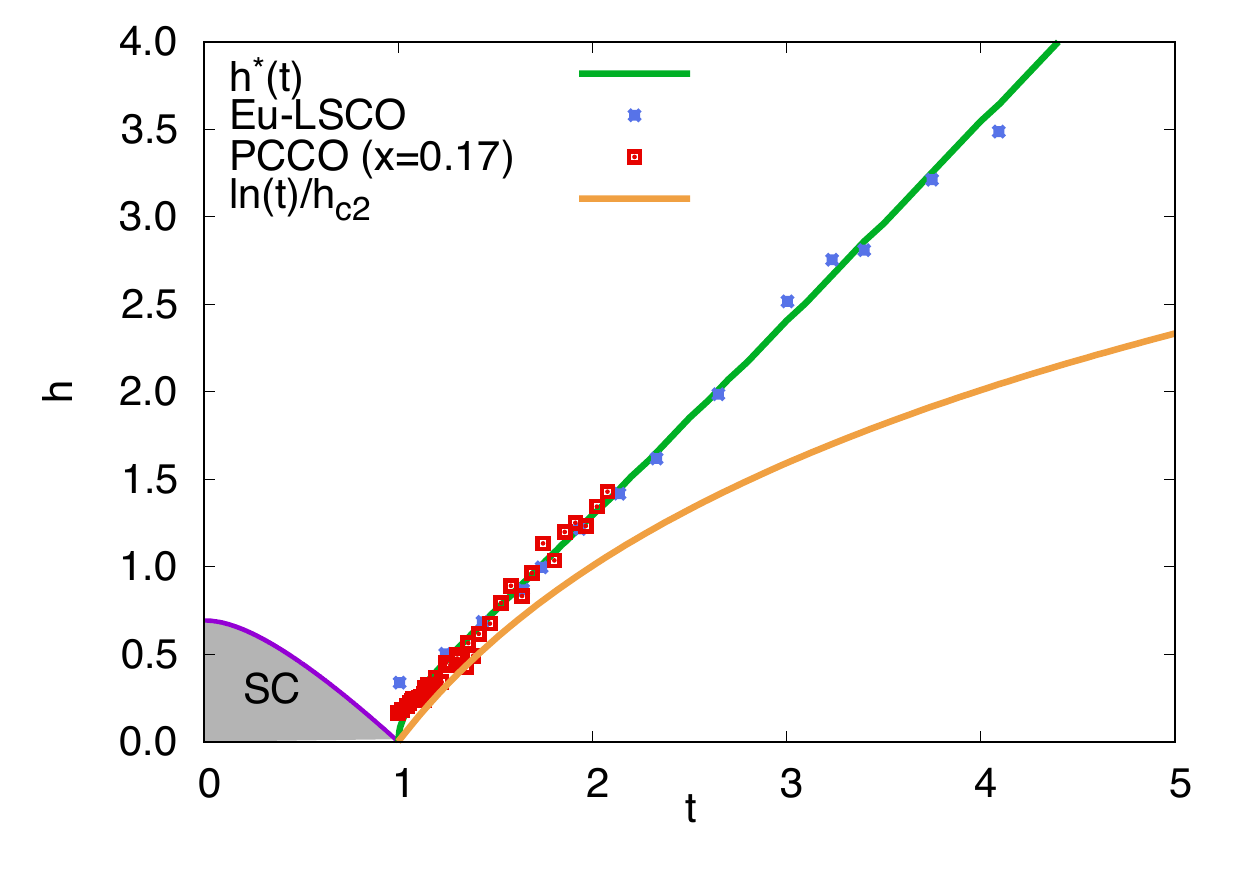} \caption{Fit of Eu-LSCO and PCCO to the numerically evaluated ghost field curve	$h^*(t)$ from Eq.~\eqref{eq:complete} compared to logarithmic dependence.}
	\label{fig:taillefer} 
\end{figure}
In Fig.~\ref{fig:taillefer}, the temperature dependence of the normalized ghost field (scaled by $\Hct(0)$) from two different experiments (dots and crosses) is compared to the numerically obtained ghost field line from Eq.~\eqref{eq:complete} (solid red line), and to the empirical $\sim\ln(t)$ (thin gray line). 
The experimental data on Eu-LSCO (purple dots) are taken from Ref.~\onlinecite{Taillefer} (Fig.~3b) and the data on PCCO at doping level $x=0.17$ (overdoped sample, green crosses) from Ref.~\onlinecite{Tafti14} (Figure~10).
One sees that the experimental findings fit the theoretical curve very well, and, in particular follow the linear behavior given by Eq.~\eqref{eq:GF_linfit}.

\begin{figure}[htb]
	\includegraphics[width=\linewidth]{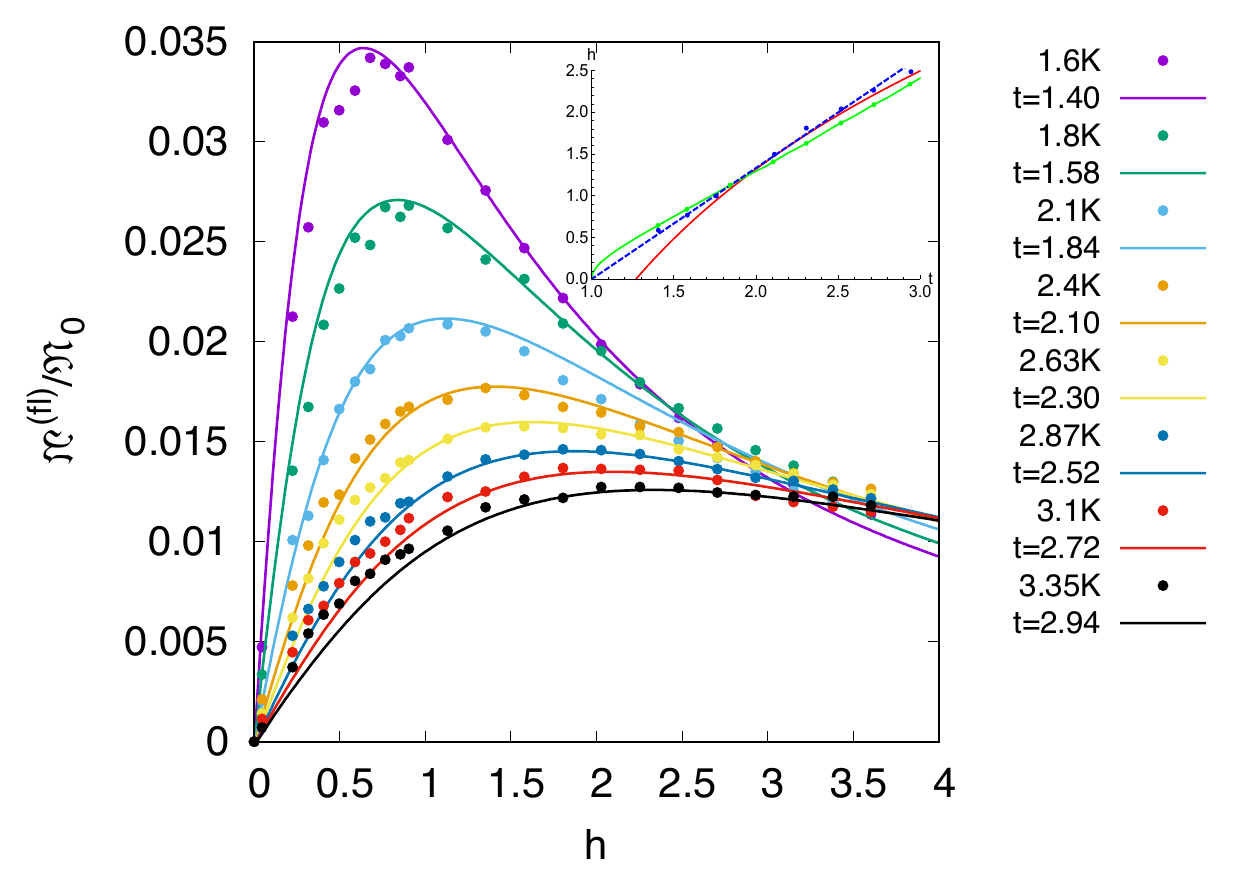} \caption{Fit of the normalized Nernst signal  vs. magnetic field measurements on heavy-Fermion superconductor URu$_2$Si$_2$~\cite{Yamashita15}
		to Eq.~\eqref{eq:complete} for different temperatures. The fitting parameters are $\Tc=1.14$K and $\Hct(0)=1.11$T.
		\textit{Insert:} the  ghost field measurements (blue circles) compared to the universal curve following from Eq.~\eqref{eq:complete} (green) and the logarithmic [Eq.~\eqref{eq:ghosttail}] fitting (red). In addition we added a linear fit to the experimental data (dashed blue). }
	\label{fig:heavy_fermion} 
\end{figure}

Finally, we also applied the numerical Nernst fitting procedure to the heavy-Fermion superconductor URu$_2$Si$_2$~\cite{Yamashita15}, where we used the measured Nernst signal data at different temperatures (Fig.~4 in that reference) and fitted $\N^{(\mathrm{fl})}\left(h\right)$ with fitting parameters $\Tc=1.14 \, $K, which is slightly lower than the empirically determined value of $1.45\, K$, and $\Hct(0)=1.11 \,$T which is close to the values found in previous experimental works~\cite{Kwok90,Moshchalkov88,Knetsch92}. 
The result is shown in Fig.~\ref{fig:heavy_fermion}.\footnote{Note that the Nernst signal curve measured at $2.1 \, K$ is labeled by $2.0 \, K$ in Fig.~4 of Ref.~[\onlinecite{Yamashita15}].}
Based on these Nernst signal fittings, we extracted the positions of maxima (the values of ghost fields) and compared them to the experimentally extracted values in the inset of Fig.~\ref{fig:heavy_fermion}.~\footnote{Our fitting procedure revealed a discrepancy in the temperature values given in Fig.~4 of Ref.~\onlinecite{Yamashita15}. We contacted one of the authors, who confirmed that $T=2.00$K should be in fact $T=2.10$K.}

\section{Discussion}
We presented a complete analysis of the magnetic field and temperature dependence of the fluctuation induced Nernst signal in a large variety of superconductors ranging from conventional to unconventional ones. 
A complete expression of the fluctuation contribution to the Nernst signal is obtained in the whole range of temperature and magnetic field and applied to experimental data by numerical analysis. 
Both magnetic field and temperature dependence of the Nernst signal data is fitted with very good accuracy using only two fitting parameters: the superconducting temperature $\Tc$ and the upper critical field $\Hct(0)$. 

Our approach predicts a linear temperature dependence for the ghost critical field well above $\Tc$, contrary to previous heuristic arguments resulting in a logarithmic dependence on temperature ~\cite{Tafti14}. 
Within the errors of the experimental data, this linearity is indeed observed in many superconductors far from $\Tc$. 
From a technical point of view we note, that the maxima of the Nernst signal become very shallow at large temperatures, which makes their extraction from experimental data very difficult. 
Therefore the seemingly simple approach to determination of the critical temperature $\Tc$ and critical field $\Hct(0)$ from the fitting of the ghost field should be done with care, giving high temperature points lower weight.

\section*{Acknowledgements}
We thank Kamran Behnia and  Maksym Serbyn for valuable discussions.
This work was supported by the U.S. Department of Energy, Office of Science, Basic Energy Sciences, Materials Sciences and Engineering Division.
A.A.V. acknowledges hospitality of Argonne National Laboratory, Northern Illinois University, and support by  European Union's Horizon 2020 research and innovation program under the grant agreement n 731976 (MAGENTA). 

\bibliographystyle{apsrev4-1}

%

\end{document}